\documentclass[fleqn,usenatbib]{mnras}

\usepackage[T1]{fontenc}
\usepackage{ae,aecompl}
\usepackage{soul}
\usepackage{mathptmx} 

\usepackage{graphicx}	
\usepackage{amsmath}	
\usepackage{amssymb}	
\usepackage{xspace} 

\def\bea{\begin{eqnarray}}
\def\eea{\end{eqnarray}}
\def\be{\begin{equation}}
\def\ee{\end{equation}}
\def\fr{\frac}

\def\ci{\cite}
\def\la{\label}
\def\be{\begin{equation}}
\def\ee{\end{equation}}

\def\le{\left}
\def\ri{\right}

\def\ac{a_c}
\def\LCDM{$\Lambda$CDM\xspace}
\def\LCDMx{$\Lambda$CDMex\xspace}

\def\nbodykit{\textsc{n-bodykit}\xspace}
\def\nbody{$N$-body\xspace}
\newcommand{\Planck}{\textit{Planck}\xspace}

\newcommand{\web}[1]{\href{#1}{#1}}

\newcommand{\hmcodelink}{\web{https://github.com/alexander-mead/HMcode}\xspace}
\newcommand{\nbodykitlink}{\web{https://nbodykit.readthedocs.io}}

\def\ad{a_{\star}}

\def\rex{\rho_{ex} }
\def\Oex{\Omega_{ex} }

\newcommand{\ve}[1]{{\text{\bf #1}}} 
\newcommand{\vk}{\ve k}

\newcommand{\vhn}{\hat{\ve n}}

\newcommand{\eg}{e.g.,\xspace}

\newcommand{\Mpc}{\,h^{-1}\mathrm{Mpc}}
\newcommand{\iMpc}{\,h\mathrm{Mpc}^{-1}}

\newcommand{\ngenic}{\textsc{ngenic}\xspace}
\newcommand{\gadget}{\textsc{gadget-2}\xspace}
\newcommand{\hmcode}{\textsc{hmcode}\xspace}

\newcommand{\ngenicaddress}{https://www.h-its.org/2014/11/05/ngenic-code/}
\newcommand{\ngeniclink}{\href{\ngenicaddress}{\ngenicaddress}\xspace}

\title[Non-linear evolution in bump cosmologies]{Impact of cosmological signatures in two-point statistics beyond the linear regime}

\author[]{
D. V. Gomez-Navarro$^{1}$,
A. J. Mead$^{2}$,
A. Aviles$^{3,4}$ \thanks{E-mail: alejandro.aviles.conacyt@inin.gob.mx}
and
A. de la Macorra$^{1,2}$
\\
$^{1}$ Instituto de F\'isica, Universidad Nacional Aut\'onoma de M\'exico,
Cd. de M\'exico C.P. 04510, M\'exico\\
$^{2}$Institut de Ci\`encies del Cosmos, Universitat de Barcelona, Mart\'i Franqu\`es 1, E08028 Barcelona, Spain\\
$^{3}$Consejo Nacional de Ciencia y Tecnolog\'ia, Av. Insurgentes Sur 1582,
Colonia Cr\'edito Constructor, Del. Benito Jurez, 03940, Ciudad de M\'exico, M\'exico. \\
$^{4}$Departamento de F\'isica, Instituto Nacional de Investigaciones Nucleares,
Apartado Postal 18-1027, Col. Escand\'on, Ciudad de M\'exico,11801, M\'exico. 
           }

\pubyear{2020}
\pagerange{\pageref{firstpage}--\pageref{lastpage}}

\begin{document}
\label{firstpage}
\maketitle

\begin{abstract}
Some beyond \LCDM cosmological models have dark-sector energy densities that suffer phase transitions. Fluctuations entering the horizon during such a transition can receive enhancements that ultimately show up as a distinctive bump in the power spectrum relative to a model with no phase transition. In this work, we study the non-linear evolution of such signatures in the matter power spectrum and correlation function using \nbody simulations, perturbation theory and \hmcode - a halo-model based method. We focus on modelling the response, computed as the ratio of statistics between a model containing a bump and one without it, rather than in the statistics themselves. Instead of working with a specific theoretical model, we inject a parametric family of Gaussian bumps into otherwise standard \LCDM spectra. We find that even when the primordial bump is located at linear scales, non-linearities tend to produce a second bump at smaller scales. This effect is understood within the halo model due to a more efficient halo formation. In redshift space these nonlinear signatures are partially erased because of the damping along the line-of-sight direction produced by non-coherent motions of particles at small scales. In configuration space, the bump modulates the correlation function reflecting as oscillations in the response, as it is clear in linear Eulerian theory; however, they become damped because large scale coherent flows have some tendency to occupy regions more depleted of particles. This mechanism is explained within Lagrangian Perturbation Theory and well captured by our simulations.
\end{abstract}


\section{Introduction}\label{Introduction}

An understanding of the Universe within the context of the \LCDM model is by now well established. High-precision measurements, such as those of the CMB from \Planck \citep{Aghanim:2018eyx} and large-scale structure in SDSS-IV \ci{Ahumada:2019vht}, including  Baryon Acoustic Oscillations (BAO) and Lyman-Alpha forest observations, can individually be well understood within \LCDM. However, there exist some observational problems within the model: there is a well-known tension in the estimated value of the rate expansion of the Universe, H$_0$, between early- and late-time observations \citep[\eg][]{Verde:2019ivm}, which may indicate the need to extend the \LCDM model; and weak gravitational lensing studies \citep[\eg][]{Abbott2018,Hikage2019,Heymans2020} seem to prefer a lower fluctuation amplitude than would be expected from CMB observations. Furthermore, from a theoretical perspective, there is no comprehensive understanding of the nature of dark matter and dark energy that \LCDM requires to together account for $\simeq95$ per cent of the energy density of the Universe today.

Extensions of the standard model of particle physics, such as in grand-unification theories \citep{pdg}, may account for dark matter and dark energy. For example, the recently proposed Bound Dark Energy (BDE) cosmology \citep{delaMacorra:2018zbk, Almaraz:2018fhb} introduces a dark gauge group SU(3) similar to the strong QCD force in the standard model. The particles in this gauge group are massless and their cosmic abundance decays like radiation at early times. However, the underlying non-perturbative dynamics cause a phase transition to take place, at a scale factor $a_c$, and the elementary particles form massive bound states (equivalent to mesons and baryons in the standard model of particle physics) and the lightest scalar field $\phi$ corresponds to dark energy. The energy density of BDE dilutes as $\rho \sim a^{-6}$ at the phase transition scale factor $a_c$ and remains subdominant for a long period of time. Eventually BDE reappears dynamically close to present time, thus accelerating the expansion of the universe. The transition leaves imprints on large-scale structure statistics \cite{Almaraz:2019zxy} -- for a model independent analysis see \cite{Jaber:2019opg} -- and a signature is generated at a scale $k_T$ entering the horizon about the phase-transition time, which shows up as a `bump' in the matter power spectrum. We call models that generate features through phase transitions Rapid Diluted Energy Density (RDED) \cite{bumps}, and BDE is one such example. Locating unusual features in the power spectrum may help in elucidating the nature of dark energy, dark matter or physics beyond the standard model. 

In this work we are motivated by RDED effects in the linear matter power spectrum. However, we will work in a model-independent way by introducing a parametrized bump to the linear matter power spectrum that will vary in position and width. We study the signatures imprinted in the linear matter power spectrum and we follow them beyond the linear regime using different, but complementary, tools.

Higher-order Perturbation theory \citep[\eg][]{Bernardeau_2002} (PT) approaches can successfully describe the intermediate, quasi-linear scales of the matter clustering, with different perturbative schemes having different advantages. 
For instance, Lagrangian Perturbation Theory (LPT) is very accurate in modelling the two-point correlation function, particularly the smearing and shift of the Baryon Acoustic Oscillation (BAO) peak, 
which although located at a large scale ($\sim 100 \,\Mpc$), is not well captured by the linear theory. On the other hand, Standard (Eulerian) Perturbation Theory (SPT) is more successful in describing 
the broadband power spectrum, but poorly models the BAO \citep[\eg][]{Tassev:2013rta,Baldauf_2015}. As soon as non-linearities become dominant, field expansions become meaningless and perturbative approaches break down. 
At this stage, the dynamics of the self-gravitating dark matter system can be tracked accurately by \nbody simulations, although these have the disadvantage of being computationally expensive. T
he non-linear dark matter statistics, however, can be described via so-called halo models \citep[\eg][]{Seljak2000, Peacock2000, Cooray2002}.

How to accurately model the power spectrum for standard \LCDM cosmologies, with standard linear spectra, has been studied in detail: PT provides high accuracy at large scales or high redshifts \cite{Bernardeau_2002}, whereas at more deeply non-linear scales one can either use fitting functions \citep[\eg][]{Smith2003, Takahashi2012}, halo-model based methods \citep[\eg \hmcode:][]{Mead2015,Mead2016,Mead20202} or emulators \citep[\eg][]{Lawrence2010, Lawrence2017}, all of which have been tuned to reproduce the power spectra measured in accurate, high-resolution simulations. Recently, attention has been focused on modelling the power spectrum `response', which is the ratio of two power spectra, with the numerator typically the cosmology of interest and the denominator typically a cosmology whose power spectrum is well known. The response has the virtue of being both easier to simulate, requiring only moderate resolution simulations, and easier to model. It has been shown that the response can be accurately modelled for dark-energy cosmologies \citep{Casarini2016, Mead2017}, modified gravity \citep{Cataneo2019}, massive neutrinos \citep{Cataneo2020} and even for the effects of baryonic feedback \citep{Mead2020}. Once an accurate model for the power-spectrum response has been developed, this can simply be converted into an accurate model for the power spectrum by multiplying by an accurate model for the power spectrum of the cosmology that appeared in the denominator when creating the response. This could be from a high-resolution simulation, fitting function, \hmcode or an emulator.

In this work we study the non-linear behavior of signatures in two-point statistics that can be approximated by a bump in the linear power spectrum, with our bump cosmology parametrized as in  equation~\eqref{linear-ps-bumpmodel}. 
We are mostly interested in the response of the bump cosmology compared to a standard \LCDM cosmology with no bump. Henceforth, our response functions are constructed simply by taking the ratios of non-linear statistics of 
cosmologies with bumps to a standard $\Lambda$CDM cosmology with no bump. We use complementary approaches to model non-linearities, for the real space power spectrum, we use 1-loop SPT,\footnote{Recently, it has been shown that features in 
the power spectrum can be indeed well modeled within PT \citep{Chen:2020ckc}.} the \hmcode model\footnote\hmcodelink, 
and low-resolution \nbody simulations. We further consider the effect of redshift space distortions in the power spectrum using the TNS model \citep{Taruya:2010mx} and the model of \cite{Scoccimarro:2004tg}. 
The non-linear correlation function is obtained through the Convolution-LPT (CLPT) of  \cite{Carlson_2012,Vlah_2015_LFFT}.

This paper is organized as follows. In Section \ref{features-power-spectrum} we review how cosmic phase transitions lead to different cosmological features; this section can be safely skipped by the reader interested in the more phenomenological aspects of our work. In Section \ref{modelling-power-spectrum} we introduce the parametric bump cosmology to be used in the rest of the work and we present specifications of our \nbody simulations suite employed to test the analytical methods. We also review different analytical models to describe the redshift and real space matter power spectrum, as well as the correlation function. In Section \ref{analysis-results}, we present the numerical and analytical results for the response functions. Finally we conclude in Section \ref{summary-conclusions}.

\section{Features in the Power Spectrum: bumps}\label{features-power-spectrum}

Models beyond \LCDM may leave different detectable features in the  matter power spectrum $P(k,a)$. These features can have different origins. For instance, in modifications of  General Relativity, as in Hordenski models \ci{Bellini:2015xja,Pogosian:2016pwr,Bayarsaikhan:2020jww, Shi:2019hwe} or  changes in the evolution of the energy density, sound speed $c_s$ and anisotropic stress of the dark matter or dark energy \ci{Linton:2017ged, de_Putter_2010, Koivisto:2005mm, Garcia-Arroyo:2020iou, Mastache:2019gui, Devi:2019hhd, Almaraz:2019zxy},as well having non-adiabatic terms which may arise if we have different rest frames of these dark fluids. Besides the effect of structure formation due to the dark sector,  the  amount of baryonic matter 
\cite{Genova-Santos:2015kma, Griffiths:2001zg}, deviations in the primordial power spectrum $P_s$ \cite{Kumazaki:2011eb} and particle content beyond \LCDM plays a significant role in the determination of the matter power spectrum.

Deviations from \LCDM cosmologies impact the evolution of the energy density evolution $\rho(t,x)$ at both the level of the background and level of the density perturbations. The background evolution has a direct effect on cosmological distances, changing for example the acoustic scale $r_s(\ad)$ or angular distances $D_A(a)$ or the growth of structure $\delta \rho(t,x)$ leaving important signatures in the  matter power spectrum as for example in \cite{Perenon:2015sla, Cadavid:2015iya, Batista:2013oca, Lee:2014maa, GilMarin:2012nb}.

Another possibility considers the addition of extra radiation at early epochs, before recombination, to reconcile the sound horizon size with the current value of the Hubble constant. In this type of extension, as a consequence of the additional extra relativistic species, the expansion rate and the growth of matter over-densities are affected, leaving detectable signals in the matter power spectrum.  
%
This extra density, henceforth $\rex$, may be considered in a certain context as extra radiation or have the generic name early dark energy (EDE) \ci{Linder:2008nq, Calabrese:2011hg}. However, there is no precise definition of EDE models and they may differ significantly on the evolution of the equation of state (EoS) $w_{ex}=p_{ex}/\rho_{ex}$. For example, earlier efforts of EDE had an EoS close to $w=-1$ at present time but significant  deviations from $w= -1$ for low redshift, e.g. $z=O(1-10)$ \ci{Bartelmann:2005fc, Grossi:2008xh, Fontanot:2012jd, Francis:2008md}, while more recently EDE models have $w = -1$  for  $z < 1000$  \ci{Klypin:2020tud,Poulin:2018cxd} similar to in the BDE model  \ci{delaMacorra:2018zbk}.


\subsection{Rapid Diluted Energy Density (RDED)}

We are interested in the cosmological consequences, and particularly imprints on the matter power spectrum, generated by an extra energy density $\rex$, beyond the standard \LCDM, for $a\leq a_c$, which dilutes rapidly after a transition takes place at the scale factor $\ac$, and with a mode $k_c=a_cH(a_c)$ crossing the horizon at such a time. For definiteness, we will assume that such component, henceforth called Rapid Diluted Energy Density (RDED)  \cite{bumps}, tracks the leading background energy density, in either radiation, matter or DE domination epochs and we study these different cases. 

Let us consider a \LCDM cosmology with and additional energy density $\rex$, and cosmic abundance 
\bea
\Oex \equiv \fr{\rex}{\rho_{sm}+\rex} = 1 - \fr{H^2_{sm}(a)}{H^2_{smx}(a)},
\la{oex}\eea
where $H^2_{sm}$ and $H^2_{smx}$ are the Hubble parameters for the model with and without the RDED component, respectively.We consider the case for which $\Oex$ is a constant for $a\leq \ac$, i.e. $\rex$ tracks the leading background energy density with $w_{ex}=w_{sm}$ at the time of the transition ($w_{sm}=1/3$ in radiation and $w_{sm}=0$ in matter domination) and a rapid dilution of $\rex$ takes place at $\ac$. At this time the EoS  $w_{ex}\equiv p_{ex}/\rex$ suffers a transition from $w_c\equiv w_{ex}(\ac)=w_{sm}$ for  $a\leq \ac$ to  $w_f \equiv w(a_f)> w_c$ for $a_f > \ac$. The value of  $\Delta w \equiv  w_f  -w_c > 0 $  and  the width $\Delta a\equiv (a_f - a_c)/a_c$ set the steepness of the transition of the energy density of $\rex(a_c) $, i.e. how fast it dilutes. For $a>\ac$  we have
\bea
\frac{\rex}{\rho_{sm}} = \le(\fr{a}{a_o}\ri)^{- \Delta w},
\eea
with $\Delta w \equiv w_f -w_c $ and for  $w_c=1/3, w_f=1$ we have $\Delta w=2/3$ the amount of extra energy density  $\Oex$ will decrease. Notice that we would expect a massive particle to go from being relativistic for $T/m \gg 1$ with $w_c=1/3$ to non-relativistic $T/m \ll 1$ with  $w_f=0$ giving a negative $\Delta w = w_f -w_c = -1/3 $ and an increasing  $\Oex$.

This property is not accomplished by an RDED component, hence it is beyond the standard model. However, it can be implemented using a scalar field $\phi$, with energy density   $\rho_\phi =E_k  + V$ and pressure $p_\phi= E_k - V$, where $E_k$ is the kinetic term and $V$ the potential energy (for a review see \citealt{Copeland:2006wr}). The evolution of the homogeneous scalar field $\phi (t)$  is given by the Klein Gordon equation
\bea \label{eq:klein_gordon} 
\ddot{\phi} + 3H\dot{\phi}+\tfrac{dV}{d\phi} = 0.
\eea

Widely used scalar potentials are either exponential or inverse power laws \ci{Steinhardt:1999nw,delaMacorra:2001xx}, e.g.
 \bea
 V(\phi)=\Lambda^4  e^{-\alpha \phi}, \;\;\;\;\   V=\Lambda^{4+n}  \phi^{-n}
 \eea
with $n$ a dimensionless constant  while  $\Lambda, \alpha^{-1}$ have mass dimension. These potentials $V$ have different behaviours depending on the value of the parameters and the initial conditions of the scalar field. The energy density $\rho_\phi$ can track the background  evolution (named tracker fields \ci{Steinhardt:1999nw,ArmendarizPicon:2000ah}) either in radiation or matter domination with $w_i=0$ or $w_i =1/3 $ and later leap to $w_f=1$  (for $V\gg E_k$) for a long period of time, diluting $\rho_\phi$ and generating a rapid dilution $\rex$ situation. 

For example, for inverse power low (IPL) scalar field potential $V=\Lambda^{4+n} \phi^{-n}$  \ci{delaMacorra:2001xx}  the evolution of $\phi$ goes through an epoch of kinetic-term domination with $w=1$, before transitioning to potential-term domination region with $w=-1$ to finally grow to a value $w\approx - 0.9$  close to present time for $n\sim 1$ \ci{Chevallier2001,Linder:2002et} and \ci{delaMacorra:2001xx,delaMacorra:1999ff} while BDE model has $n=2/3$ \ci{delaMacorra:2018zbk}. On the other hand a value of $n>5$ has a tracking behaviour, i.e. the scalar field abundance evolves as the dominant background energy density, but these models have an EoS $w>- 0.7$ and are ruled out by current cosmological observations \cite{Aghanim:2018eyx}. Models with $n<2$ are viable as DE models. 

An example of an IPL model is BDE, which is derived from particle physics where the fundamental particles are massless and contained in a gauge group similar to the strong QCD force, however due to the underlying dynamics they develop a IPL potential with $n=2/3$ at the transition with a scale factor $a_c\sim 10^{-6}$. In this case the EoS is $w_c=1/3$ at early times and at $a_c$ it leads to $w_f=1$ to a later evolve to $w\sim -1$ at $z\sim 1000$ and finally ending up at a value of $w\sim - 0.93$  at present time. The complete evolution of $w(\phi)$ is a consequence of the dynamics of the equation of motion and is not set by hand. The BDE example satisfies the condition ($\Delta w = w_f -w_c >0$) of RDED twice.  
The first is during radiation domination at $a_c \sim 10^{-6}$ with a $k_c=0.94$ and the second takes place close to present time ($z\sim 0.28$). The transitions we have been studying here are clearly beyond the standard model of particle physics but are expected in phase transitions for dynamical scalar fields and are motivated to explain the nature of dark energy.

\subsection{Evolution of matter perturbations in RDED models}\label{implications_deltam}

Matter perturbations are affected by RDED, leaving distinctive features in cosmological observables such as the matter perturbations. The amplitude of these perturbations is increased in RDED cosmology compared to a standard \LCDM cosmology, and a bump feature 
is produced in the ratio of power spectrum $P_{smx}/P_{sm}$. The bump is generated for modes  $k > k_c$, with $k_c\equiv a_cH(a_c)$, entering the horizon at a scale factor
$a < a_c$.

We will sketch how this bump is generated and we will study in detail the properties of these bumps in both the linear and non-linear approach. Let us explore here the linear regime of matter density fluctuations. In this case, the evolution of CDM perturbations obey the equation of motion:
\bea\label{eq:implications_deltac}
 \delta_c''+ \mathcal{H} \delta_c'-\frac{3}{2} \mathcal{H}^2\sum_i \Omega_i \delta_i (3c_{s,i}^2+1)=0\ ,
\eea
where the sum runs over all the fluids with sound speed $c_{s,i}^2=\delta P_i/\delta \rho_i$ and the prime denotes derivatives with respect to conformal time, with $\mathcal{H}=a'/a$. We see that the $\rex$ has a twofold effect on matter perturbations by modifying first the expansion rate $\mathcal{H}$, and second by the adding of an extra source term proportional to $\Omega_{ex}\delta_{ex}$.

Initially, for modes outside the horizon, the matter perturbations do not evolve over time, leaving the ratio $Q_m\equiv \delta_m^\textrm{smx}/\delta_m^\textrm{sm}$ fixed. However, matter perturbations in the model with $\rex$ are initially suppressed compared to a \LCDM model because the initial amplitude of the matter perturbations, through the gravitational potential $\Psi$, depends on the fraction of  relativistic particles $R_\nu = \rho_{eff}/(\rho_{eff}+\rho_{\gamma})$, with $\rho_\gamma$ the density of photons, $\rho_{eff}^{\textrm{sm}}=\rho_\nu$  and $\rho_{eff}^\textrm{smx}=\rho_\nu + \rex$  giving an initial suppression factor $Q_{ini}=(1+(4/15)R_\nu^{\textrm{sm}})/(1+(4/15)R_\nu^{\textrm{smx}})$, which depends solely on $\rex$.
  
Modes start evolving once they cross the horizon $a_h$, defined implicitly in terms of the Hubble radius by $k=a_h H(a_h)$. However, there is a marked difference depending on whether they cross before $a_c$, with a corresponding mode $k_c\equiv a_cH(a_c)$, or after the extra energy density $\rex$ has diluted. Small modes $k>k_c$ are further suppressed with respect to \LCDM because the crossing time in this case is delayed by the presence of the $\rex$ .
Comparing the same $k_h=(a_h H(a_h))\big|_{smx}=(a_h H(a_h))\big|_{sm}$ we have
\be\label{eq:implications_ah}
 \frac{a_h^{\mathrm{smx}}}{a_h^\mathrm{sm}}=\frac{H^{sm}}{H^{smx}}=\fr{1}{\sqrt{1-\Oex}}\ ,
\ee
and therefore small modes $k>k_c$ cross the horizon earlier in $\Lambda$CDM  than  in \LCDMx.
This is reflected in an early suppression in  $\Delta \delta_m =\delta \rho_m^{smx}/\delta\rho_m^{sm}$, supplemented by a smaller amplitude at horizon crossing. 
However, after the initial suppression at horizon crossing, matter perturbations in the $smx$  model have a higher growing rate than in the $\Lambda$CDM model that not only compensates but also reverses the initial suppression \citep{bumps}.

In radiation domination the matter perturbations evolve as $\delta_m \propto \ln a$ with the growth function 
$f= d\ln\delta_m /d\ln a \propto 1/\delta_m$,
and since  $\delta_m(smx) < \delta_m(sm)$ and $f(smx)> f(sm)$ and $Q_m$ increases. 
This increase is boosted by the rapid dilution of $\rex$, where the EoS leaps abruptly from   $w_i=1/3$ to $w_f=1$ at $a_c$ affecting only the modes crossing the horizon before $a_c$, i.e. $k \geq  k_c$.  This is the characteristic signature of RDED and was presented in \ci{Almaraz:2018fhb} and \ci{Jaber:2019opg}. However, to fully asses the growth of structure in these RDED cosmologies we must go beyond the linear regime, which is the main goal of this work.

To gain physical intuition on how the rapid diluted energy density affects matter-density fluctuations well inside the radiation dominated epoch we can analyze a simplified version of the equations where the gravitational potentials are subdominant: We can estimate this increase in radiation domination, where the amplitude $\delta_m=\delta \rho_m/\rho_m$ has a logarithmic growth
\be\la{deltama}
\delta_m (a)  = \delta_{mi}  \le[\textrm{ln}(a/a_h)+1/2 \ri]\ ,
\ee
with $a_h$ and $H_h$ the scale factor and Hubble parameter at horizon crossing. We compare the evolution of the same mode $k^{smx}=k^{sm}$
crossing the horizon before $a_c$ and using equations~(\ref{deltama}) and (\ref{eq:implications_ah}) we find the ratio 
$\Delta \delta_m  = 
\delta_m^{smx}/\delta_m^{sm} =
(\delta_{mi}^{smx}/\delta_{mi}^{sm})([\ln(a/a_h^{smx})+1/2]/[\ln(a/a_h^{sm})+1/2])$.
The evolution of $\Delta \delta_m$ after the RDED transition takes place at $a_c$ is given by \ci{bumps}
\be
\Delta \delta_m = \fr{ \delta_{mi}^{smx}}{\delta_{mi}^{sm}}\;
 \fr {\le[  \le(\fr{H_+^{smx}}{H_-^{smx}}\ri) \textrm{ln} \le(\fr{a}{a_c}\ri) +\textrm{ln} \le( \fr{ a_h^{sm}}{a_h^{smx}}\ri) + \textrm{ln} \le( \fr{a_c}{a_h^{sm}}\ri)+\fr{1}{2}   \ri] }
{ \textrm{ln}\le(\fr{a}{a_c}\ri) + \textrm{ln}\le(\fr{a_c}{a_h^{sm}}\ri) +\fr{1}{2} }
\la{DD}\ ,
\ee
where we have taken into account in $\Lambda$CDMx  the contribution of $\rex$ in $H_+^{smx}(a_c)$ for $a<a_c$ and  $\rex=0$ in $H_-^{smx}(a_c)$ for  $a> \ac$, giving the ratio $H_-^{smx}/H_+^{smx}= a_h^{\mathrm{smx}}/a_h^\mathrm{sm}=\sqrt{1-\Oex}$  (c.f. equation~(\ref{eq:implications_ah})). The matter power spectrum therefore shows a bump for modes $\Delta \delta_m$ for $ a\gg a_c$ for modes $k>k_c$ entering the horizon before $a_c$ when we compare RDED and \LCDM cosmologies.



\section{Modelling the matter power spectrum} \label{modelling-power-spectrum}

To encompass a range of theoretical models, we choose to work with a parametrization that we refer throughout as the `bump cosmology', where the linear power spectrum is given by a modification to that of a standard \LCDM cosmology:
\begin{equation}
    P_\text{bump}(k, z)=\big[1+ F(k)\big]P_\text{$\Lambda$CDM}(k, z)\ ,
    \label{linear-ps-bumpmodel}
\end{equation}
with
\begin{equation}
F(k)=A \exp \left(-\frac{[\ln(k/k_\mathrm{T})]^2}{\sigma^2} \right)\ .
\label{bumpmodel}
\end{equation}
$A$, $k_T$ and $\sigma$ are the amplitude, scale, and width of the bump, respectively. We considered other choices for the function $F(k)$, such as a Gaussian in $k$-space, and found qualitatively similar results.

We consider nine different bump cosmologies, in each case we fix the amplitude $A=0.15$. 
We choose this amplitude motivated by BDE, where the energy density of the dark sector $\Omega_\mathrm{BDE}(a_c)=0.11$  dilutes  rapidly at  $a_c=2.48\times 10^{-6}$ due to a
phase transition of the underlying physics and rendering  a $\Omega_\mathrm{BDE} \ll 1$ 
at scales $a > a_c$.  Meanwhile, the width of the bump corresponds to how fast the rapid diluted energy density takes place, for which we investigate three different values: $\sigma=1.0$, $0.3$, and $0.1$, and place the bump at three different scales: $k_T=0.05$, $0.1$, and $1\iMpc$ (see Table \ref{tab:simulation}). We study structure formation in these bump cosmologies at the redshifts $z=0$, $0.5$, $1$, $2$, $3$, and $4$. The cosmological parameters used to generate the \LCDM power spectrum are reported below in Section \ref{subsect:NbodySims} and are the same in both the bump and standard cosmologies, so that the \emph{only} difference between the models is the presence of the bump.


We investigate the response of the real-space matter power spectra and correlation functions together with the redshift-space multipole power spectra. We construct the response as the ratio of the measurement or prediction between a bump and \LCDM cosmology. We consider different approaches for the non-linear theory: moderate-resolution \nbody simulations, and \hmcode. In addition, for the real-space power spectra we also consider one-loop SPT, for the redshift-space multipole power spectra we use the TNS and Scoccimarro models and finally, for the real-space correlation function we use CLPT.

\subsection{N-body simulations}
\label{subsect:NbodySims}

\begin{table*}
	\centering
	\begin{tabular}{l c c c c} 
		\hline
		\\ [-3pt]
		Name & $\qquad A \qquad $ & $\qquad \sigma \qquad $ & $k_T \,\,[\iMpc ]$ & $L_{box} \, \,[\Mpc ]$ \\ [2pt]
		\hline
		\\ [-3pt]
		\textsc{fatbump-k1} & $0.15$ & $1.0$ & $1.0$ & 256\\[4pt]
		\textsc{medbump-k1} & $0.15$ & $0.3$ & $1.0$ & 256\\[4pt]
		\textsc{thinbump-k1} & $0.15$ & $0.1$ & $1.0$ & 256\\[4pt]
		\textsc{$\Lambda$CDM-k1} & $-$ & $-$ & $-$ & 256\\[4pt]
		\textsc{fatbump-k0p1} & $0.15$ & $1.0$ & $0.1$ & 512\\[4pt]
		\textsc{medbump-k0p1} & $0.15$ & $0.3$ & $0.1$ & 512\\[4pt]
		\textsc{thinbump-k0p1} & $0.15$ & $0.1$ & $0.1$ & 512\\[4pt]
		\textsc{$\Lambda$CDM-k0p1} & $-$ & $-$ & $-$ & 512\\[4pt]
		\textsc{fatbump-k0p05} & $0.15$ & $1.0$ & $0.05$ & 1024\\[4pt]
		\textsc{medbump-k0p05} & $0.15$ & $0.3$ & $0.05$ & 1024\\[4pt]
		\textsc{thinbump-k0p05} & $0.15$ & $0.1$ & $0.05$ & 1024\\[4pt]
		\textsc{$\Lambda$CDM-k0p05} & $-$ & $-$ & $-$ & 1024\\[4pt]
		\hline
	\end{tabular}
	\caption{Specifications of our \nbody simulation suite. 
	The background cosmological parameters are the same for all the simulations: $\Omega_m = 0.3$, $\Omega_b = 0.05$, $\Omega_{\Lambda} = 0.7$, $\Omega_{\nu} = 0$, $h = 0.7$, $n_s = 0.96$, $\sigma_8 = 0.8$. Each simulation uses $512^3$ particles distributed over $N_{grid}=512^3$ cells. We consider the redshifts $z=0,0.5,1,2,3,4$.}
	\label{tab:simulation}	
\end{table*}

We ran $12$ \nbody simulations using the cosmological simulation code \gadget\citep{Springel2005}, one each for the cosmologies detailed in Table~\ref{tab:simulation}. We assume a background \LCDM cosmology with total matter density $\Omega_m = 0.3$, baryon density $\Omega_b = 0.05$, dark energy density $\Omega_{\Lambda} = 0.7$, amplitude of the matter power spectrum $\sigma_8 = 0.8$, spectral index $n_s = 0.96$, and dimensionless Hubble constant $h = 0.7$. Initial conditions were generated at $z=99$ using \ngenic,\footnote\ngeniclink which implements the Zeldovich approximation to displace particles from an initially Cartesian grid and assigns them velocities based on a ballistic trajectory. We chose to run simulations with different box sizes for the different $k_T$ values, to ensure that there was always a good sampling of modes around $k_T$. The box sizes of the simulations are $L_{box} = 256$, $512$ and $1024\Mpc$, for $k_T=0.05$, $0.1$ and $1\iMpc$ respectively. Each simulation uses $512^3$ particles to approximate the density field, which is quite modest compared to modern simulation standards. One may worry that measurements from these simulations would be systematically biased as a result of the low mass resolution. However, we checked for convergence with respect to low-resolution $256^3$ particle simulations and found that our results for the power spectrum \emph{response} were only significantly affected ($>1$ per cent) on scales smaller than half of the particle Nyquist frequency, although there were some noticeable, sub per-cent differences for scales as large as one tenth of the particle Nyquist frequency. This is different to a general suppression that affects the power spectrum (rather than the response) when using low-resolution simulations \citep{Mead2015}, and must be because some of the bias in power that arises when using a low particle number cancels when constructing a response; so the same (or a similar) bias must occur in the cosmologies of both the numerator and denominator that make up the response.

\subsection{\hmcode}

We look at predictions for the non-linear matter power spectrum of our cosmological models using the \hmcode \citep{Mead2015} model. \hmcode is an augmented version of the traditional halo-model calculation for the non-linear power spectrum \citep[\eg][]{Seljak2000, Peacock2000, Cooray2002} where the augmentations account for defects and replace ingredients that are missing from the standard halo-model calculation. This improves the accuracy of the calculation from $\sim 30$ per cent to $\sim 5$ per cent for a wide range of cosmologies and redshifts. \hmcode takes as input the linear power spectrum of the cosmology in question, and then uses some information about the background cosmological parameters and power-spectrum shape and amplitude in order to make its predictions. Although \hmcode was not calibrated on the bump cosmologies investigated in this paper, the grounding of the model in physical reality means that we can hope that it will make reasonable predictions. \hmcode models the power spectrum as a sum of (almost) linear theory and a one-halo term, which accounts for small-scale, deeply non-linear power under the assumption that all such power originates from the clustering within haloes. The bump cosmologies will therefore affect the \hmcode predictions in two ways. The first, is trivially that the \hmcode prediction at large scales is essentially linear theory and because linear theory contains the bump then so will \hmcode. The second, is that within \hmcode the one-halo power is determined by the halo mass function, which itself depends on the linear power spectrum via the variance in the power spectrum as a function of scale. The bump will therefore affect the halo mass function and we expect that it will boost the predicted numbers of haloes in certain mass ranges.

Before running \hmcode, we can make the prediction that it should generally boost power in the one-halo term for a bump cosmology compared to a cosmological model that lacks a bump. The traditional halo model calculation has a problem in the transition region between the two- and one-halo terms ($k\sim 0.05\iMpc$ at $z=0$), and generically underestimates the true non-linear power spectrum in this region; this probably arises due to an improper treatment of non-linear halo bias \citep[][]{Smith2007}. The solution to this problem in \hmcode is a smoothing of the transition region. Based on this discussion, we could predict that the \hmcode predictions for the bump cosmologies may be better in the deeply one-halo regime ($k>1\iMpc$), and that they may be less impressive in the transition region. In the linear region they should be perfect, given that the \hmcode prediction is identical to linear theory at large scales. In the quasi-linear regime ($k\sim 0.1\iMpc$ at $z=0$) we would expect perturbation theory to perform better than \hmcode since the latter lacks any formal grounding in analytical perturbation theory.

\subsection{Standard perturbation theory}

In this paper we consider two variations of perturbation theory, the first, `standard' perturbation theory is constructed in Eulerian space, where we consider the evolution of the density field at fixed positions. We consider the phenomenological, but physically well motivated, redshift-space models of \citet{Scoccimarro:2004tg} and \citet{Taruya:2010mx} to describe the redshift-space power spectrum, which improve the description of redshift-space distortion (RSD) effects on the power spectrum compared to Kaiser linear theory \cite{Kaiser:1987.10.1093/mnras/227.1.1}. In the following we shall refer to these models as Sc04 and TNS, respectively. 

The anisotropic redshift-space clustering originates from the peculiar velocities $\mathbf{v}$ of matter, or more generally any tracer of it, such that an object located at a real space position $\mathbf{r}$ is observed to be located at an apparent redshift-space position $\mathbf{s}$. The relation between coordinates system is inferred via the Doppler effect to be $\mathbf{s}=\mathbf{r} + \hat{\mathbf{n}} v_{\parallel} (aH)^{-1}$, where $\hat{\mathbf{n}}$ is a the line-of-sight direction of the point-process sample, and $v_{\parallel}=\mathbf{v} \cdot \hat{\mathbf{n}}$. That is, we are using the plane-parallel approximation on which the observer is located at a distant position of the sample of objects over which we perform the statistics.
The  redshift-space power spectrum, $\langle |\delta^s(\mathbf{k})|^2 \rangle$ is given by
\begin{align}    \label{redshift-space-power}
    P^s(\mathbf{k})&=\int d^3r \, e^{i\mathbf{k}\cdot\mathbf{r}} \Big\langle e^{ik\mu\Delta v_\parallel/(aH)}\left (\delta(\mathbf{x})-\frac{1}{aH} \nabla_\parallel v_\parallel(\mathbf{x}) \right)  \nonumber \\
    &\quad \qquad \times \left (\delta(\mathbf{x'})-\frac{1}{aH} \nabla_\parallel v_\parallel(\mathbf{x'}) \right) \Big\rangle 
\end{align}
where $\mathbf{r}=\mathbf{x}-\mathbf{x'}$ and $\Delta v_\parallel =v_\parallel(\mathbf{x})-v_\parallel(\mathbf{x'})$ and  $\mu=\hat{\mathbf{n}}\cdot \hat{\mathbf{k}}$ is the angle between the wave vector and the line-of-sight direction.  

The RSD correction at linear order, known as the Kaiser formula, is given by $\delta^s_L(\mathbf{k})=(1+f\mu^2)\delta_L(\mathbf{k})$, where $f=d \log D_+(a)/d \log a(t)$ and $D_+(t)$ the linear growth function. The redshift-space power spectrum at linear order becomes 
\begin{equation}
    P^{s}_K(k, \mu)=(1+f\mu^2)^2 P_L(k).
    \label{kaiser-equation}
\end{equation}

To move beyond linear theory it is common to define the dimensionless velocity divergence $\theta= -\nabla {\bf \cdot v}/(aHf)$, such that at linear order we have $\theta=\delta$.

The exponential oscillatory factor inside the correlator in equation~\eqref{redshift-space-power} is due to virialized, non-coherent random motions of dark-matter particles along the line-of-sight direction, hence it is in essence non-perturbative. In \citep{Scoccimarro:2004tg} this factor is replaced by a phenomenological damping function that accounts for the velocity dispersion $\sigma^2_v=\langle \theta^2 \rangle$. By rotational symmetry around the line-of-sight direction one obtains a simple prescription
\begin{align} \label{Sc04-equation}
P^s_{\text{Sc04}}(k,\mu)&=\exp\big[-k^2\mu^2f^2\sigma^2_v\big] \nonumber\\
&\qquad \times \left[ P_{\delta\delta}(k)+2f\mu^2P_{\delta \theta}(k)+ f^2\mu^4P_{\theta\theta} (k) \right],
\end{align}
where $P_{\delta\delta}$, $P_{\theta\theta}$ and $P_{\delta\theta}$ are respectively the non-linear matter density, velocity divergence, and density-velocity divergence power-spectra. In linear theory equation~\eqref{Sc04-equation} reduces to the Kaiser power spectrum (equation~\ref{kaiser-equation}) times the damping factor. 
Several other functional forms for this damping factor have been used in the literature. In this work, we opt for the most common -- Gaussian damping. 
The velocity dispersion is physically motivated to be given by PT as 
\begin{equation}
 \sigma^2_v = \frac{1}{6 \pi^2}\int dp \, p^2 P_{\theta\theta}(p),
\end{equation}
but due to its non-perturbative origin it is commonly replaced by a free parameter $\sigma^2_\text{FoG}$, especially for parameter estimation.
However in this work, we regard $\sigma^2_v$ as the linear velocity dispersion, which is obtained from the above equation by replacing $P_{\theta\theta}$ by its linear value $P_L$. This prescription has shown to give good results when comparing theory to
simulations \citep[\eg][]{Taruya:2010mx}.

The TNS formalism, on the other hand, expands in cumulants the correlator in equation~\eqref{redshift-space-power}, and then replaces a residual exponential factor of the form $\exp\big[ \langle e^{ik\mu\Delta v_\parallel/(aH)}\rangle_c \big]$, by a position independent, phenomenological damping factor that can be brought out of the integral. The standard formula for TNS is
\begin{align}
P^s_{\text{TNS}}(k,\mu)=&\exp(-k^2\mu^2f^2\sigma^2_v) \big[ P_{\delta\delta}(k)+2f\mu^2P_{\delta \theta}(k) \nonumber\\ 
&+ f^2\mu^4P_{\theta\theta}(k)  +A(k,\mu)+B(k,\mu)],
\end{align}
where the new correction terms, $A(k,\mu)$ and $B(k, \mu)$, are given by
\begin{align}
        A(k, \mu)&=2 k \mu f \int \frac{d^3p}{(2 \pi)^3} \frac{\mathbf{p}\cdot\hat{\mathbf{n}}}{p^2} B_\sigma (\mathbf{p},\mathbf{k}-\mathbf{p},-\mathbf{k}), \\
        B(k, \mu)&= (k \mu f)^2 \int \frac{d^3p}{(2 \pi)^3} F(\mathbf{p}) F(\mathbf{k}-\mathbf{p}) ,
\end{align} 
with the bispectrum
\begin{align}\label{defBsigma}
&(2\pi)^3 \delta_\text{D}(\vk_1+\vk_2+\vk_3)B_\sigma(\vk_1,\vk_2,\vk_3) = \Big\langle \theta(\vk_1) \nonumber \\
&\times \left[\delta(\vk_2) + f \frac{(\vk_2\cdot\vhn)^2}{k_2^2}\theta(\vk_2) \right] 
  \left[\delta(\vk_3) + f \frac{(\vk_3\cdot\vhn)^2}{k_3^2}\theta(\vk_3) \right]\Big\rangle,
\end{align}
and
\begin{equation}
 F(\mathbf{p}) =  \frac{\mathbf{p}\cdot\hat{\mathbf{n}}}{p^2} \left( P_{\delta\theta}(p) + f \frac{(\mathbf{p}\cdot\hat{\mathbf{n}})^2}{p^2} P_{\theta\theta}(p) \right).
\end{equation}
Hence, the TNS model partially accounts for the interaction between the Kaiser effect and the non-linear random motion of particles, reflected in the two extra functions $A$ and $B$. On the other hand, the Sc04 model factorizes the linear and Finger-of-God effects, so each can be treated separately.

The real-space matter power spectrum is given by $P_{\delta \delta}(k)=\langle |\delta (\mathbf{k})|^2 \rangle$, which can be obtained from both of the above redshift-space formalisms by considering only the perpendicular to line-of-sight components (i.e., $\mu=0$). After that, one can simply use rotational symmetry to obtain the 1-loop SPT power spectrum
\begin{align}
P_{\text{1-loop}}^{\text{SPT}}(k,t)&= P_{L}(k)+ P_{22}(k) + P_{13}(k),
\label{power-1loop-spt}
\end{align}
where  $P_{22}$ and $P_{13}$ are the usual 1-loop corrections \citep[\eg][]{Bernardeau_2002}. 
The linear power spectrum is the dominant term at large scales. However, at smaller scales, the loop corrections contribute with similar magnitude to the linear power, therefore significantly contributing to the total power spectrum.

\subsection{Lagrangian Perturbation Theory}\label{clpt-subsection}

In contrast to the Eulerian scheme, in a Lagrangian picture we follow the trajectories of individual particles with initial position $\mathbf{q}$ and current position $\mathbf{x}$. The Lagrangian coordinates $\mathbf{q}$ are related to the Eulerian coordinates $\mathbf{x}$ through the coordinates transformation
\begin{equation}
    \mathbf{x}(\mathbf{q},t)=\mathbf{q}+\Psi(\mathbf{q},t),
\end{equation}
where $\Psi(\mathbf{q},t)$ is the Lagrangian displacement and $\Psi(\mathbf{q},t=t_{ini})=0$, meaning that at a sufficiently early time $t_{ini}$ both coordinates system coincide. Furthermore, mass conservation implies the relation between Lagrangian displacements and overdensities \citep{Taylor:1996ne}
\begin{equation}
    1+\delta(\mathbf{x},t)=\int{d^3q} \delta_D[\mathbf{x}-\mathbf{q}-\Psi],
\end{equation}
from which the LPT correlation function is obtained as 
 \begin{eqnarray}\label{xilpt}
     1 + \xi(r)=\int \frac{d^3 k}{(2 \pi)^3} \int d^3q e^{i\mathbf{k}\cdot (\mathbf{r}-\mathbf{q})}\langle e^{-i\mathbf{k} \cdot \Delta} \rangle,
 \end{eqnarray}
 where $\Delta^i=\Psi^i(\mathbf{q}_2)-\Psi^i(\mathbf{q}_1)$ are the Lagrangian displacement differences at two positions $\mathbf{q}_1$ and $\mathbf{q}_2$ separated by a distance $q=|\mathbf{q}_2-\mathbf{q}_1|$. The idea behind CLPT is to expand the correlator in eq.~\eqref{xilpt} in cumulants and thereafter to expand out of the exponential all but linear terms in the Lagrangian displacement, such that the $\vk$-integral can be performed analytically using standard Gaussian integration techniques (see \citet{Carlson_2012}), by which we arrive at
\begin{align}\label{xiCLPT}
    1+\xi_{\text{CLPT}}(r)&=\int {\frac{d^3q}{(2\pi)^3 \text{det}[A_{ij}^{L}]^{1/2}}}e^{\frac{1}{2}A_{ij}^L(r_i-q_i)(r_j-q_j)} \nonumber \\
    &\quad \times \left[1-\frac{1}{2}G_{ij}A_{ij}^{\text{loop}}+\frac{1}{6}\Gamma_{ijk}W_{ijk}^{\text{loop}} + \cdots\right],
\end{align}
with cumulants $A_{ij}=\langle \Delta_i \Delta_j\rangle_c$ and $W_{ijk}=\langle \Delta_i \Delta_j \Delta_k\rangle_c$, and tensors $G_{ij}=A^{-1}_{L\,ij}-g_ig_j$, $\Gamma_{ijk}=A^{-1}_{L\,ij}g_k+A^{-1}_{L\,jk}g_i+A^{-1}_{L\,ki}g_j-g_ig_jg_k$, and $g_{ij}=A_{L\,ij}^{-1}(r_j-q_j)$. The label  `$L$' in the $A$ function denotes the linear piece and  `loop' the pure 1-loop piece, such that $A_{ij}=A^L_{ij}+A^\text{loop}_{ij}$. Notice that the `1' in the squared brackets of the above equation corresponds to the Zeldovich correlation function and the other two terms yield the next-to-leading order, 1-loop contributions. 

To solve numerically the CLPT correlation function of equation~\eqref{xiCLPT} we use the code \textsc{mgpt} \citep{Aviles:2018saf},\footnote{\href{https://github.com/cosmoinin/MGPT}{https://github.com/cosmoinin/MGPT}} which also accounts for the exact kernels for a \LCDM background cosmology, instead of the most commonly used Einstein-de Sitter ($\Omega_\mathrm{m}=1$) kernels. Using the correct kernels can make a small, but noticeable, difference to the power spectrum -- as much as $0.8$ per cent at quasi-linear scales at $z=0$.

\section{Results and analysis}\label{analysis-results}

We use the different approaches discussed in the previous Section to study the evolution, dilution and shift of the bump as seen in the response functions
\begin{equation}
    R(k) = \frac{P_\text{bump}(k)}{P_\text{$\Lambda$CDM}(k)}.
\end{equation}
That is, we use our \nbody simulations, \hmcode and PT methods as complementary tools. The results are contrasted with the linear theory, for which the response in the power spectrum is simply $R_L(k) = 1+ F(k)$ at all $z$ because linear growth is scale independent.

\subsection{Real-space matter power spectrum} 
\begin{figure*}
    \centering 
	\includegraphics[trim=2cm 1.85cm 2cm 2.8cm, clip=true, width=0.98\textwidth]{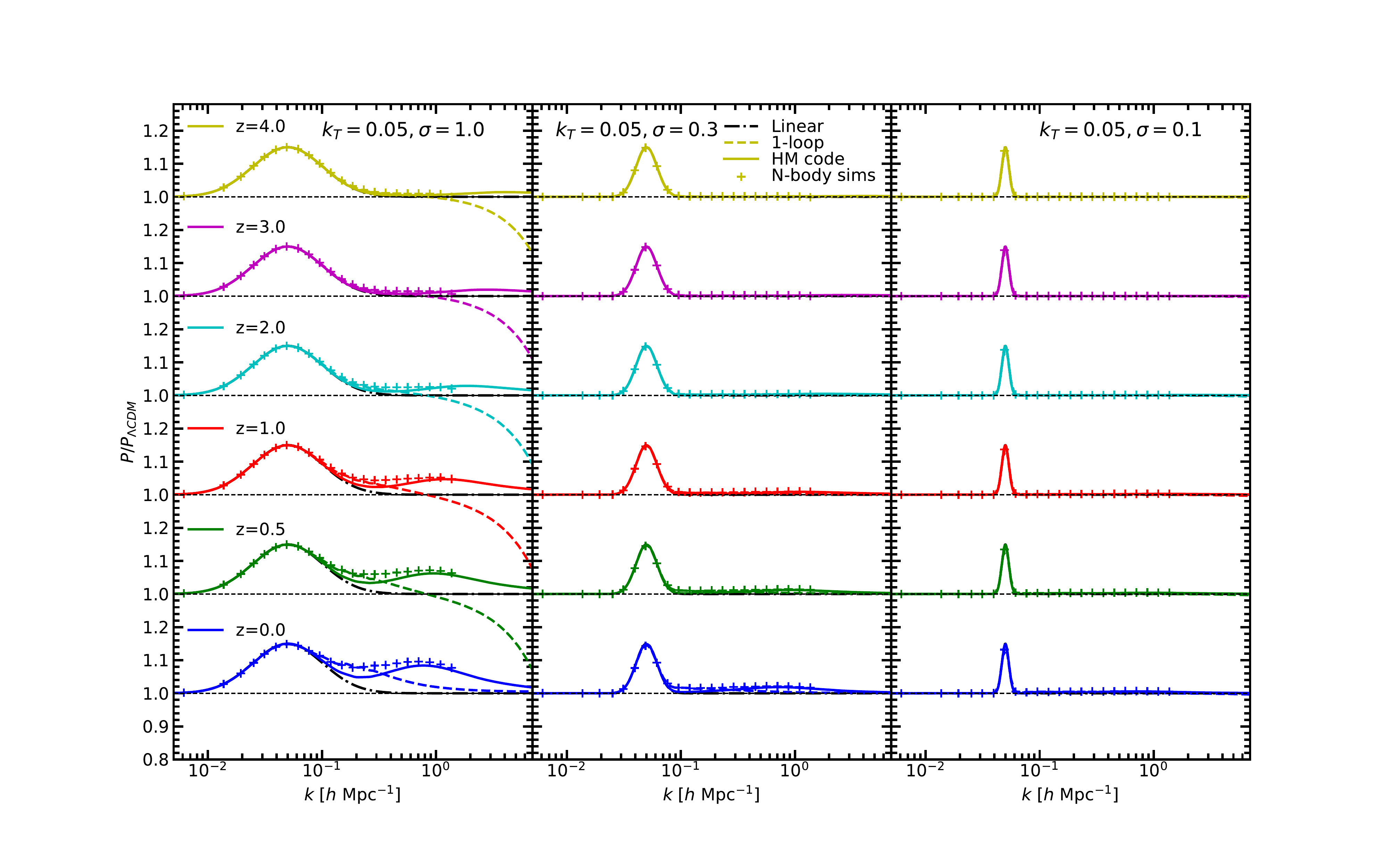} 
    \caption{Response functions for the bump cosmologies at $k_T=0.05\iMpc$. From top to bottom, yellow curves are for redshift $z=4$; magenta for $z=3$; cyan for $z=2$; red for $z=1$; green for $z=0.5$; and blue for $z=0$. The left panel shows the bump cosmology for $\sigma=1$; middle panel for $\sigma=0.3$; and right panel for $\sigma=0.1$. Dash-dotted (black) curves are for the linear theory; dashed (color) are for 1-loop SPT; solid for \hmcode model; and crosses are for the measurement from \nbody simulations.}
    \label{power-spectra-k0p05}
\end{figure*}

\begin{figure*}
    \centering 
	\includegraphics[trim=2cm 1.85cm 2cm 2.8cm, clip=true, width=0.98\textwidth]{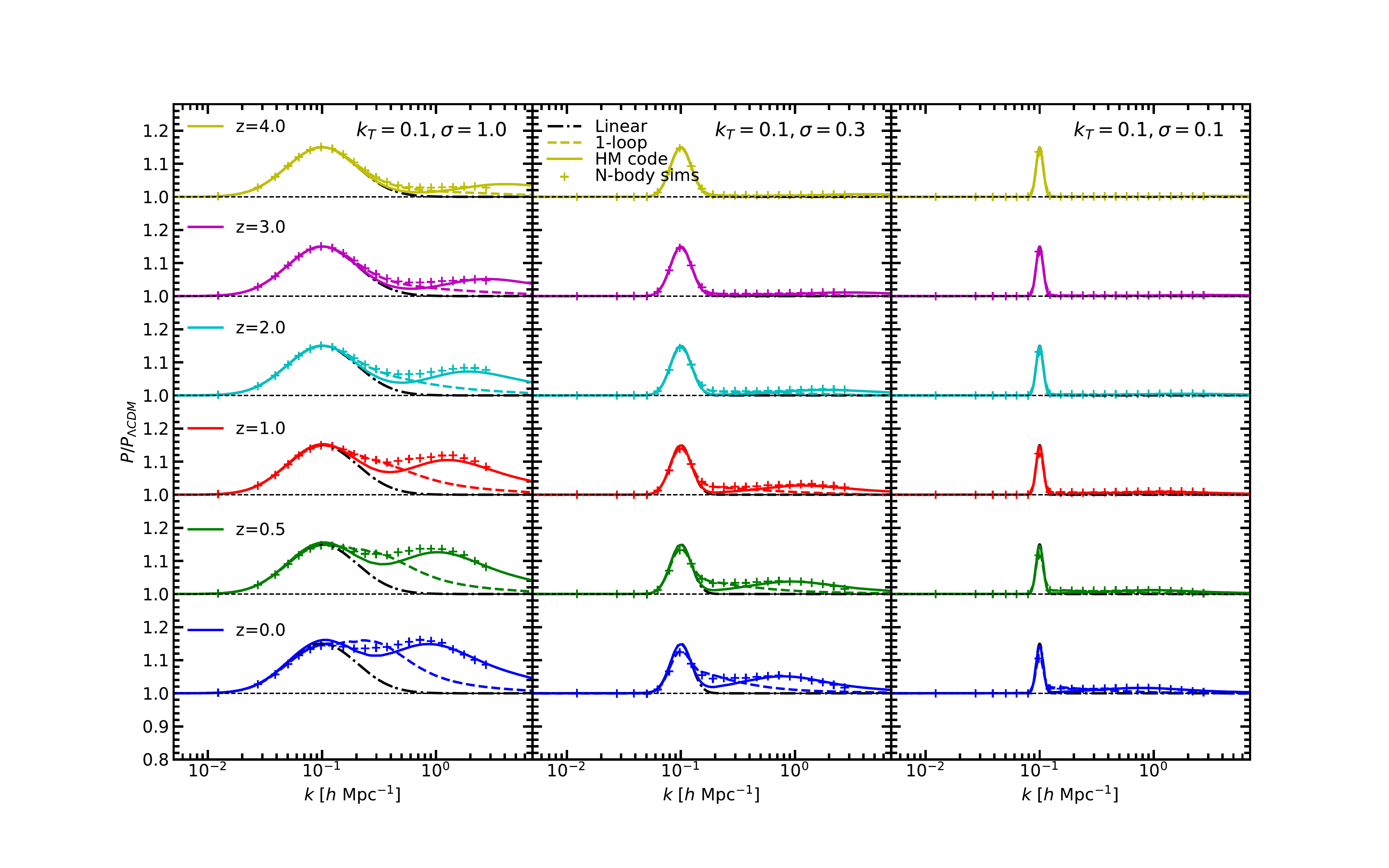} 
    \caption{Same as Fig. \ref{power-spectra-k0p05} but for the bump cosmologies with $k_T=0.1\iMpc$.}
    \label{power-spectra-k0p1}
\end{figure*}

\begin{figure*}
    \centering 
	\includegraphics[trim=2cm 1.85cm 2cm 2.8cm, clip=true, width=0.98\textwidth]{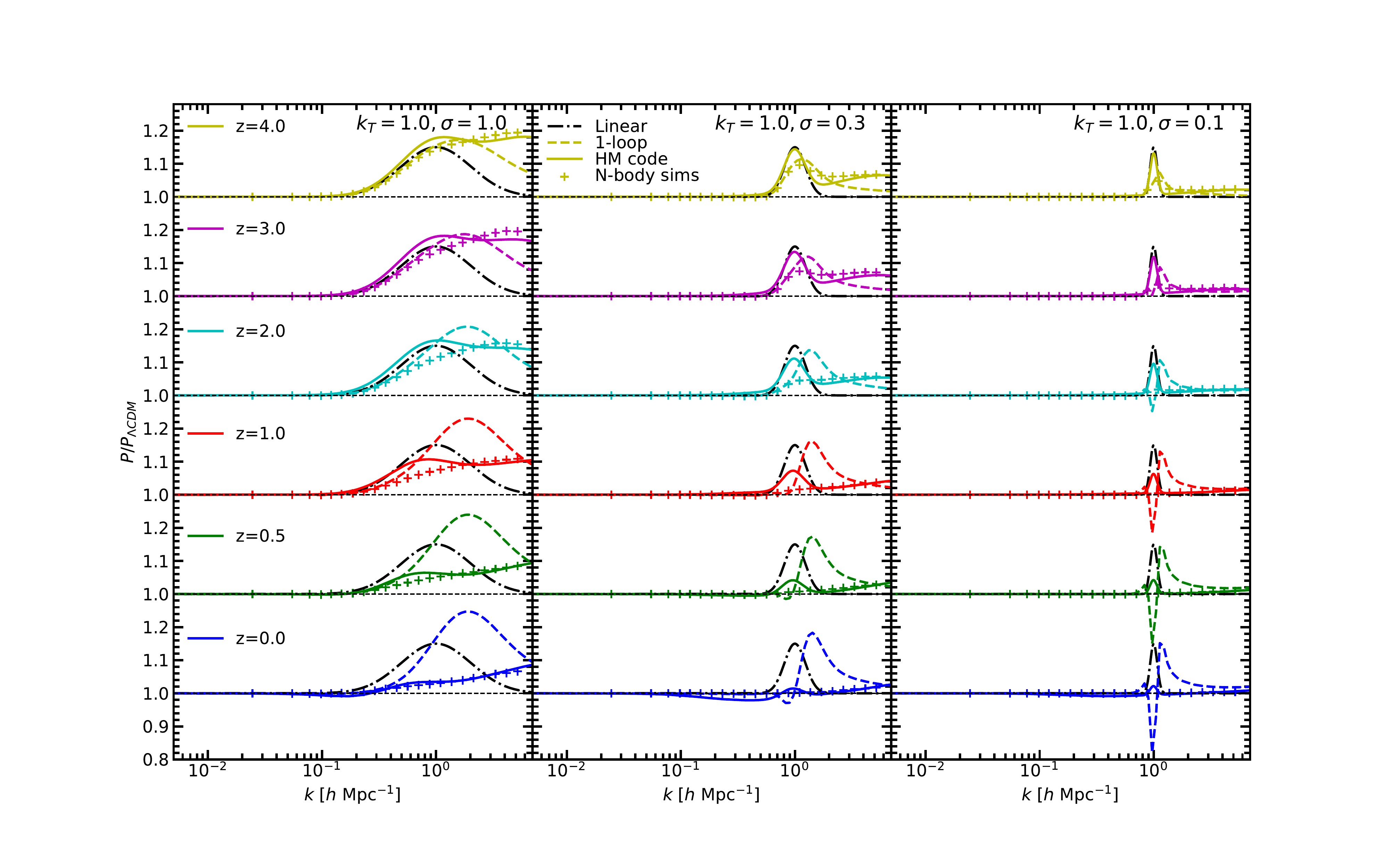} 
    \caption{Same as Fig. \ref{power-spectra-k0p05} but for the bump cosmologies with $k_T=1\iMpc$.}
    \label{power-spectra-k1}
\end{figure*}

The power spectra data are extracted from the simulations using a cloud-in-cell mass-assignment scheme, on a grid with resolution of $N_{grid}=512^3$ cells. They are binned in $100$ evenly log-spaced $k$-points over a range $[k_{min}, k_\mathrm{Ny}]$, where $k_\mathrm{Ny} = N_{grid} \pi /L$ is the Nyquist frequency, and $k_{min}= 2\pi /L$. $L$ is the size of the box, which is given in Table \ref{tab:simulation} for the different simulations. Traditionally power spectra measured in this way are considered to be accurate up to half of the Nyquist frequency, with the power at smaller scales potentially corrupted by aliasing from smaller scales. In our plots we show all measured scales, up to $k_\mathrm{Ny}$, and the reader should keep this in mind.

Figs. \ref{power-spectra-k0p05}, \ref{power-spectra-k0p1}, and \ref{power-spectra-k1} show the responses using the bump cosmology from equation~\eqref{linear-ps-bumpmodel} located at the scales $k_T=0.05$, $0.1$, and, $1\iMpc$, respectively. 
The non-linear power is computed using our different methods and then divided by their counterparts in \LCDM to create the response function. This analysis compares how the bump cosmologies are modified by non-linearities within the different approaches.

As expected, at higher redshifts non-linear effects are smaller and the responses of all methods are very similar. In the $\sigma=1$ case of Fig.~\ref{power-spectra-k0p05} as $z$ decreases we see two things in the simulated measurements: First, at the small-scale edge of the bump we see the power grow above the bump, an effect that is captured extremely well by SPT and less well by \hmcode. Second, we see the generation of a second bump at much smaller scales, with a peak $k\sim1\iMpc$. This second bump is well modelled by \hmcode but not captured at all by SPT. The failure of SPT at these scales (which even predicts a decrease in the response at higher $z$) is not surprising given that these scales are far beyond its reach. In Fig.~\ref{power-spectra-k0p05}, corresponding to $k_T=0.05 \iMpc$, this non-linear `second bump' peaks at $10$ times smaller scale, $k \approx 0.5\iMpc$, and reaches a maximum at $z=0$, peaking in amplitude at $\sim 10$, $2$, and $1$ per cent for the $\sigma=1$, $0.3$, and $0.1$ cases respectively. For $k_T=0.1$, shown in Fig. \ref{power-spectra-k0p1}, this second non-linear bump is even clearer, peaking at $k \approx 1\iMpc$ for $z\simeq 0$, but at smaller scales for higher redshifts. For some of our cosmologies, this second bump is even larger than the primordial bump. For example, it is larger for $\sigma=1$ at both redshifts $z=0.0$ and $z=0.5$, contributing $\approx 18$ and $16$ per cent to the whole response respectively. We note that as the width of primordial bump decreases, the second, non-linear bump amplitude decreases. In all cases, the location and amplitude of the non-linear bump as seen in the simulation response is in remarkable agreement with the predictions from \hmcode. For the cases of $k_T =0.05$ and $0.1\iMpc$, shown in Figs.~\ref{power-spectra-k0p05} and \ref{power-spectra-k0p1}, \hmcode and the simulated data provide similar results at lower redshifts and in the non-linear regime. Conversely, 1-loop SPT gives results closer to those obtained from simulations at higher redshifts and in the mildly non-linear regime ($k\lesssim 0.2\iMpc$).

The non-linear bumps in the response functions are a consequence of the interaction between the primordial bump and the one-halo term, being highly enhanced for the wider bumps simply because these provide a greater enhancement of linear power. Physically, this can be understood within the halo model (and therefore within \hmcode) as follows: the one-halo term is given by the integral of the halo mass function multiplied by the squared Fourier-space halo profile. The halo mass function itself is related to the standard-deviation in the density field when smoothed over the Lagrangian radius of the halo, $\sigma_R$. Our bump cosmology increases the power over a certain range of scales, such that $\sigma_R$ will also increase, and therefore so will the mass function. Hence, one effect of the linear bump is to accelerate halo formation relative to a cosmology with no bump. A different way to think about this is via a \cite{Press1974} type argument, where the increased amplitude of some modes, given here by our bump cosmology, is helping more small scale fluctuations to cross over the critical threshold to collapse. This occurs even when the bump is at very linear scales as long as the width is sufficiently large, because those long wavelength modes exist underneath the smaller-scale fluctuations enhancing the collapse to form the actual haloes. We remark that this is a highly non-linear effect, such that the second bump is not well captured by PT, where the main non-linear effect is the spreading and enhancement of the primordial bump.

In Fig.~\ref{power-spectra-k1} we show the  $k_T =1\iMpc$ bump cosmologies. The simulated data show that the primordial bump become erased by the non-linear evolution. Such effect is well captured by the \hmcode  -- PT simply does not capture the non-linearities of the evolution of the bump since it is far out of the reach of its regime of validity, although at the highest redshifts it works moderately well.
Interesting is the non presence of the second bump, because the primordial bump and the 1-halo term are located about the same scale.

Figs. \ref{power-spectra-k0p05}, \ref{power-spectra-k0p1} and \ref{power-spectra-k1} demonstrate that perturbation theory provides an accurate model for the power spectrum response at large scales, with the accuracy extending to smaller scales at higher redshift. On the other hand, \hmcode provides a reasonable (though less perfect) model for the response at the smaller scales, those typically associated with halo formation. It does not take a great leap of the imagination to consider combining these two approaches to provide an accurate model for the response that would be valid across a wider range of scales. This approach has been explored in the literature previously \citep[\eg][]{Mohammed2014a, Seljak2015, Philcox:2020rpe}, but has never been applied to the bump cosmology specifically. It may be possible to directly add perturbation theory to the halo-model, perhaps replacing the two-halo term with some perturbative expansion, or else to use the response from perturbation theory at one extreme and \hmcode at the other, interpolating between the two around some `non-linear' wavenumber, perhaps defined via $\sigma(a/k_\mathrm{nl})=\alpha$ where $\sigma$ is the standard deviation in the density field when smoothed on a given physical scale and $a$ and $\alpha$ are fitted constants of order unity.

\begin{figure*}
    \centering 
	\includegraphics[trim=2cm 0cm 2cm 1cm, clip=true, width=0.9\textwidth]{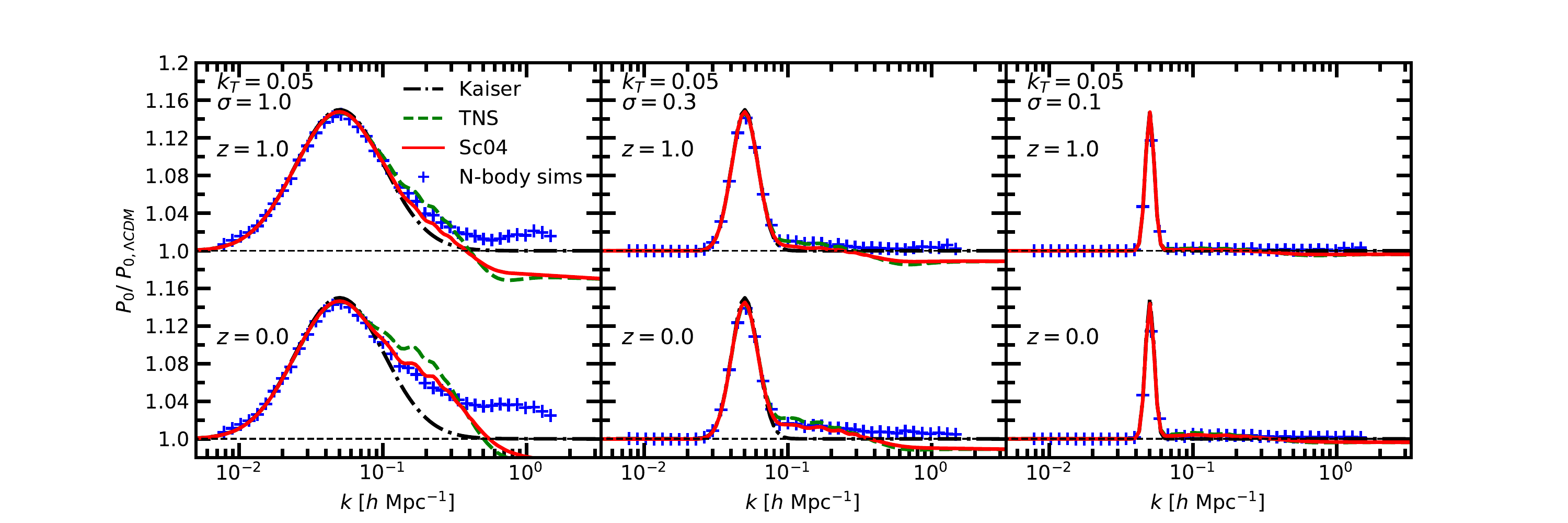}
	\includegraphics[trim=2cm 0cm 2cm 1cm, clip=true, width=0.9\textwidth]{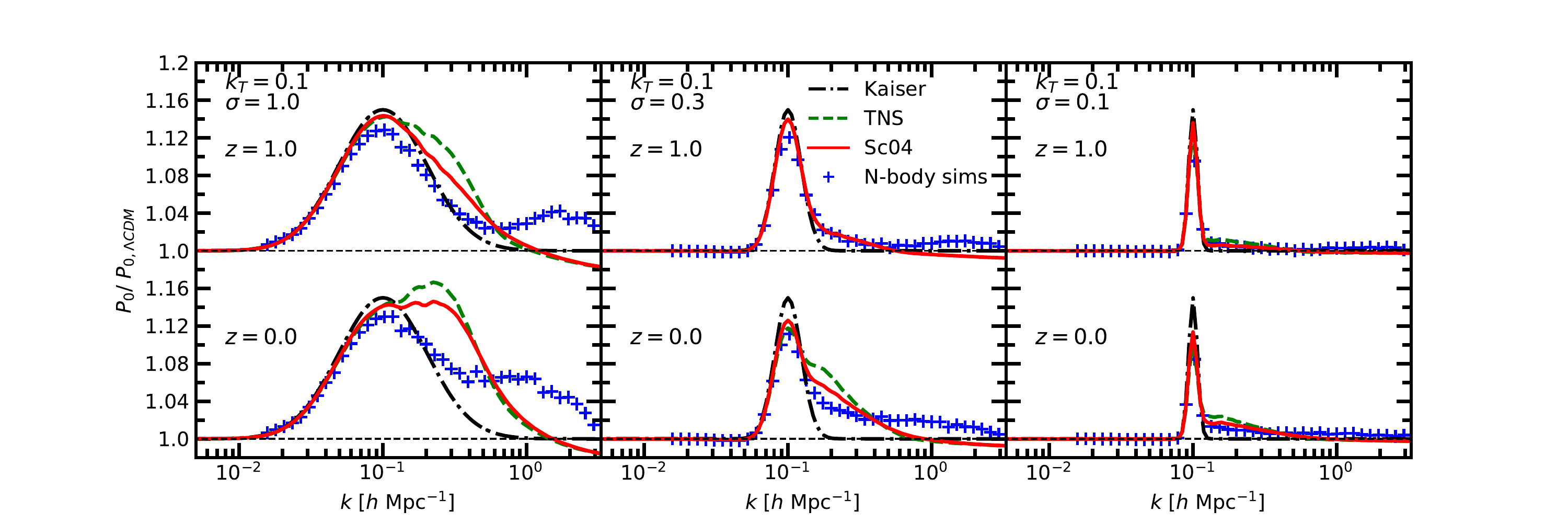}
    \caption{Responses of the monopole power spectrum at $z=0$ and $1$ for the bump cosmologies at $k_T=0.05$ (top panel) and $0.1\iMpc$ (bottom panel), with different widths $\sigma=1, 0.3, 0.1$ (from left to right). Dashed black shows the Kaiser linear theory; dashed green is for the TNS model; solid red for Sc04 model; and the blue crosses show the simulated data.} 
    \label{monopole-power-k0p05}
\end{figure*}

\begin{figure*}
    \centering 
	\includegraphics[trim=2cm 0cm 2cm 1cm, clip=true, width=0.9\textwidth]{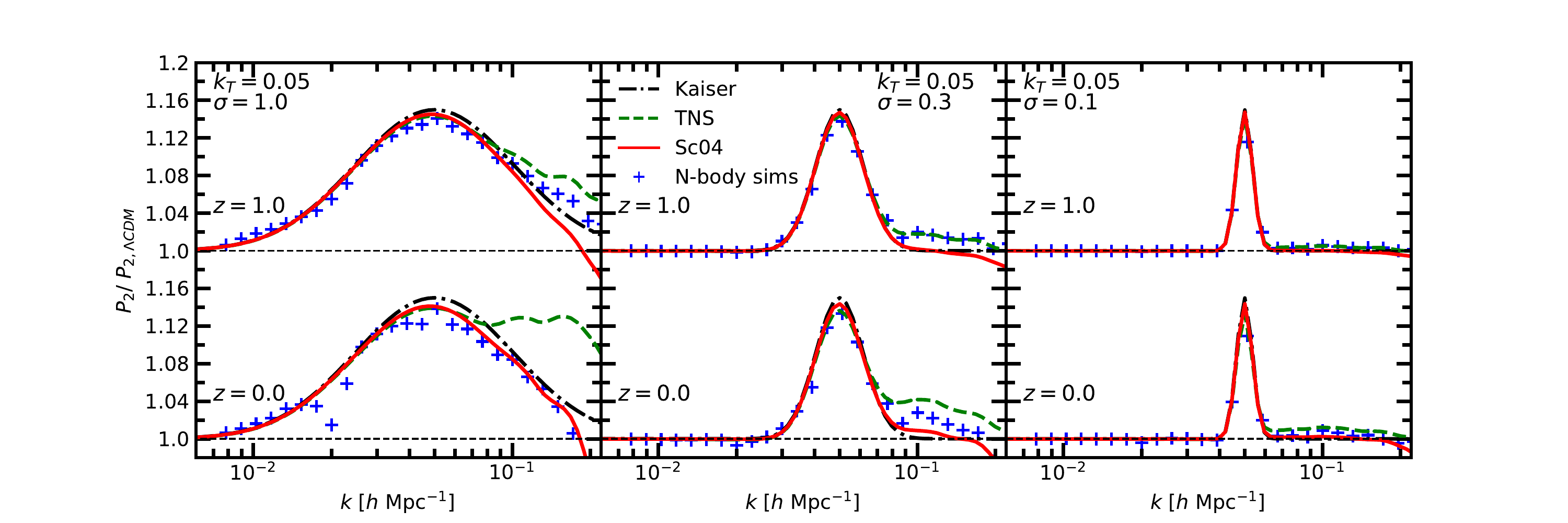}
	\includegraphics[trim=2cm 0cm 2cm 1cm, clip=true, width=0.9\textwidth]{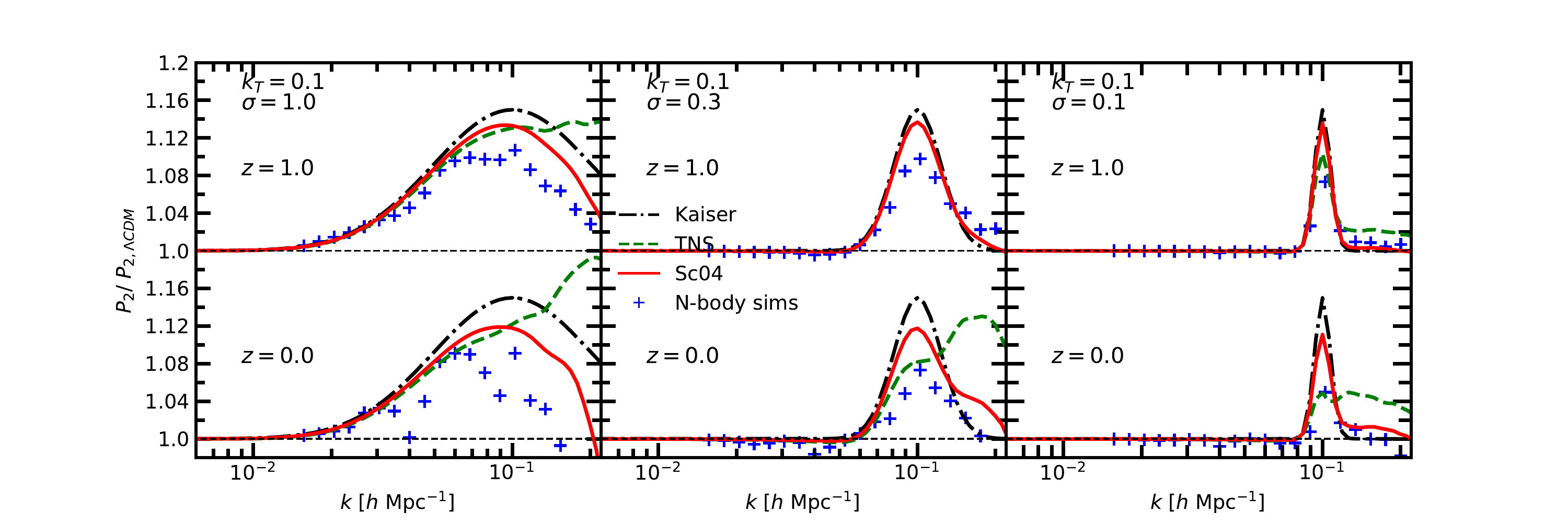}
    \caption{As Fig. \ref{monopole-power-k0p05} but for the quadrupole response.} 
    \label{quadrupole-power-k0p05}
\end{figure*}

\subsection{Redshift-space matter power spectrum multipoles}

In this Section, we study the responses of the redshift-space monopole and quadrupole power spectra using the non-linear models Sc04 and TNS. For comparison we also use Kaiser linear theory. The multipole power spectra have been extracted from the \nbody simulations using a triangular-shaped-cloud mass-assignment function, on a grid with $N_{grid}=512^3$ cells, using the \nbodykit package\footnote\nbodykitlink of \cite{Hand:2017pqn}. The response functions are defined for each multipole, such that $R_\ell(k) = P_\ell^\text{bump}(k)/P_\ell^\text{$\Lambda$CDM}(k)$.

In the top panel of Fig.~\ref{monopole-power-k0p05} we show the response functions in the monopole $P_0$ at redshifts $z=0$ and  $z=1$ for the bump cosmologies with $k_T=0.05 \iMpc$.
We notice, as expected, that non-linear models perform closer to the simulated data than the linear theory in the quasi-linear regime, in the sense that the responses are closer to those from \nbody simulations. However, for $k > 0.2 \iMpc$ both theoretical models make deviations from the simulated data. In the bottom panel of Fig.~\ref{monopole-power-k0p05} we show the response functions for the bump cosmologies located at $k_T=0.1\iMpc$. The behaviour of the PT methods is qualitatively similar to the case of $k_T=0.05 \iMpc$, but the non-linear features of the bump are less well captured when comparing to the simulated data. This must be since the scale of the bump is located at the edge of the perturbative regime of validity. This has the consequence that the linear theory response provides a better match to the simulations for some non-linear scales, obviously the good performance of linear theory here is only a lucky coincidence. As in real space, the simulated data show the appearance of the non-linear second bump at $\approx 1\iMpc$, which is not present in the $k_T=0.05 \iMpc$ case, even though in real space this second non-linear bump was visible in both cases. We suggest that the reason for this is the damping along the line-of-sight direction of the redshift-space power spectrum, which ultimately comes from the highly oscillatory behavior of the correlator inside the integral of equation~\eqref{redshift-space-power} at large $k$. This redshift-space effect damps the multipoles in all bump cosmologies, but since the second (real space) bump is larger for the case of $k_T=0.1\iMpc$, it can overcome the damping and it still shows up in the redshift-space responses. 

We also show, in Fig.~\ref{quadrupole-power-k0p05}, the quadrupole redshift-space power spectra for $k_T=0.05 \iMpc$ and $k_T=0.1 \iMpc$. Although the simulated quadrupole measurements are noisier than for the monopole, we note a similar trend to that predicted by the analytical approaches, most obviously for the case of $k_T=0.05\iMpc$ at $z=1$. We also observe that the non-linear, second bumps do not appear in the quadrupole. We suggest that this is because this multipole gives maximum weight to the line-of-sight direction where the damping effect occurs, while the monopole gives equal weight to all directions.

\subsection{Real-space matter correlation function}\label{sec:resultsCF}

\begin{figure*}
    \centering 
	\includegraphics[trim=2cm 1.85cm 2cm 2.8cm, clip=true, width=0.98\textwidth]{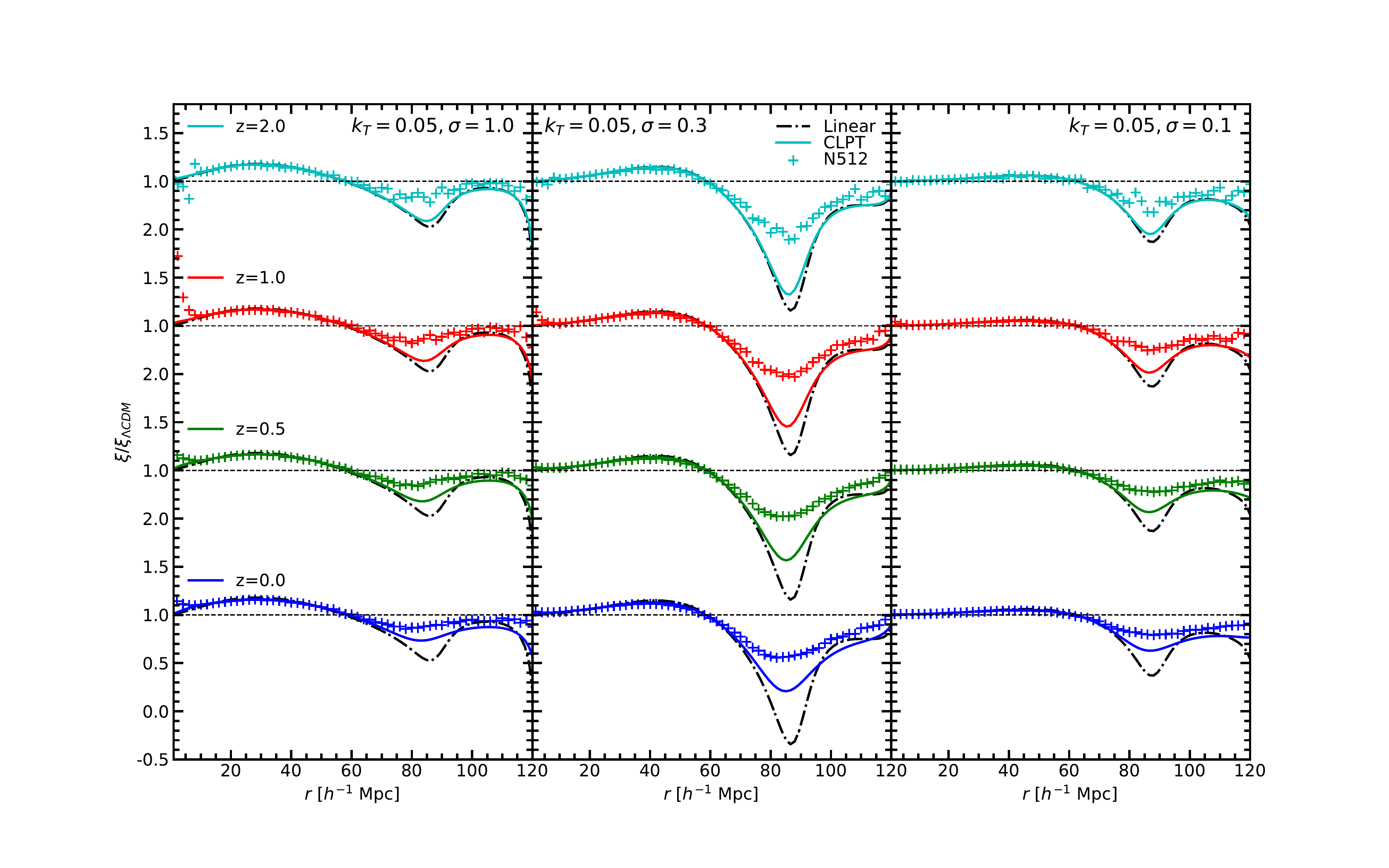} 
    \caption{The real-space correlation function response: We plot $\xi/\xi_{\Lambda CDM}$ for bump cosmologies at $k_T=0.05$ in units $\iMpc$ for different redshifts ($z=0.0, 0.5, 1.0, 2.0$). From top to bottom, cyan curves are for $z=2$; red for $z=1$; green for $z=0.5$; and blue for $z=0$. Right: Bumps with $\sigma=1.0$. Middle: Bumps with $\sigma=0.3$. Left: Bumps with $\sigma=0.1$. The CLPT correlations are represented by the solid lines. The plus markers denote the \nbody simulations. The black dash-dotted lines denote the linear correlation function.}
    \label{corr-k0p05}
\end{figure*}

\begin{figure*}
    \centering 
	\includegraphics[trim=2cm 1.85cm 2cm 2.8cm, clip=true, width=0.98\textwidth]{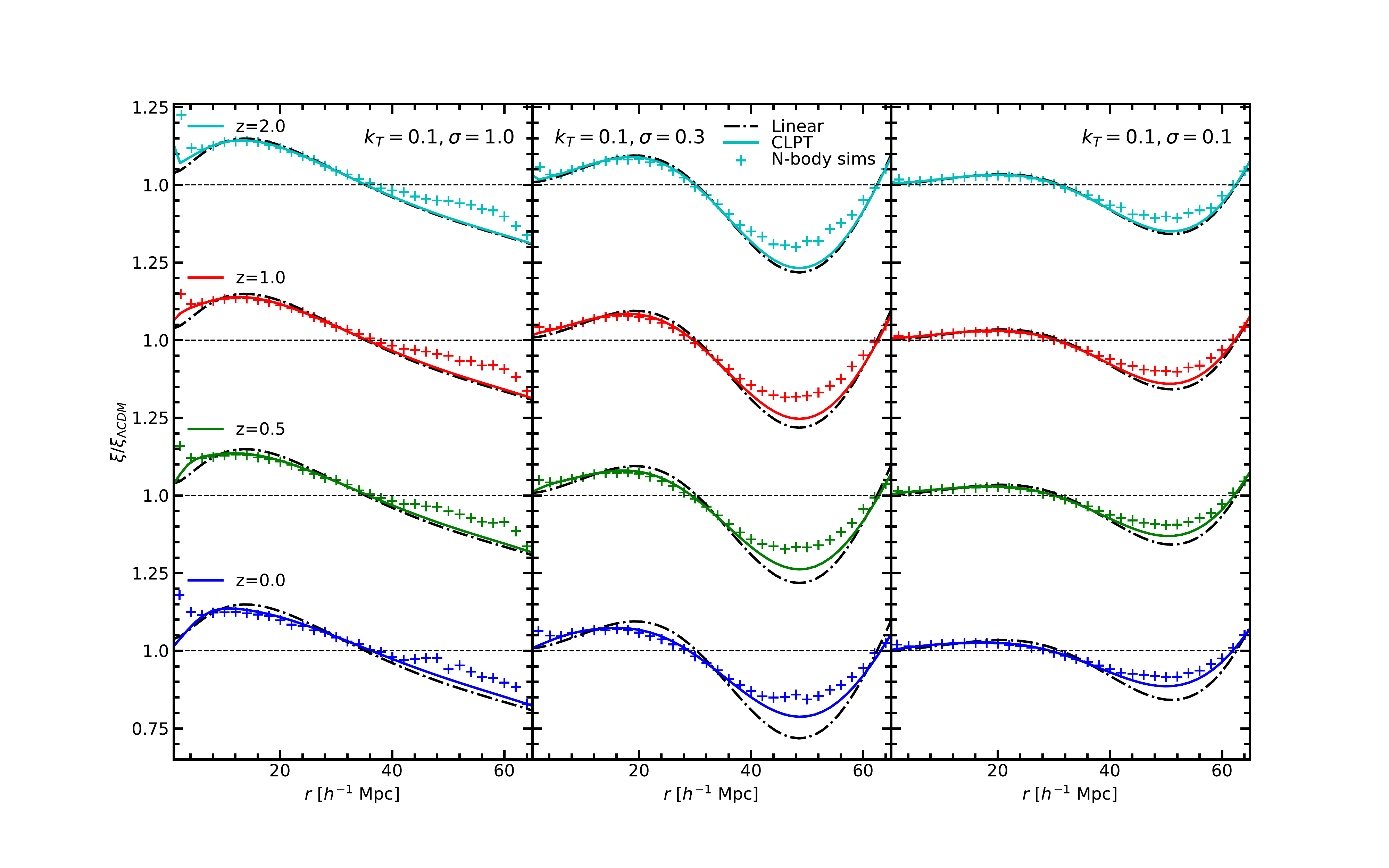} 
    \caption{As Fig. \ref{corr-k0p05} but for the bump cosmologies with $k_T=0.1\iMpc$.} 
    \label{corr-k0p1}
\end{figure*}

\begin{figure*}
    \centering 
	\includegraphics[trim=2cm 1.85cm 2cm 2.8cm, clip=true, width=0.98\textwidth]{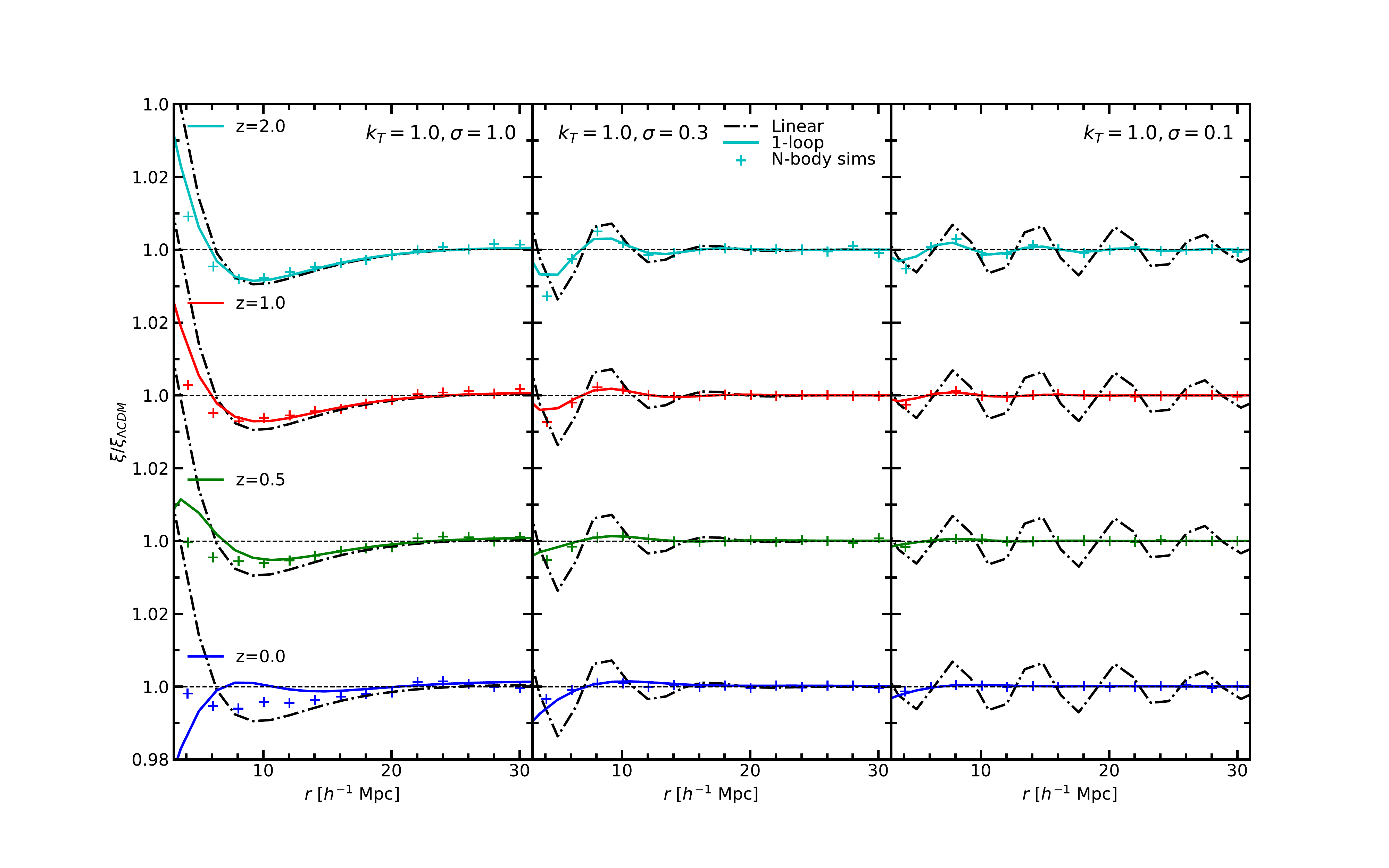} 
    \caption{As Fig.~\ref{power-spectra-k0p05} but for the bump cosmologies with $k_T=1\iMpc$.} 
    \label{corr-k1}
\end{figure*}

In this sub-section we compare the measured real-space correlation function responses, defined as $\xi_\text{bump}(r)/ \xi_\text{$\Lambda$CDM}(r)$, from both the simulated data and the  analytical correlation function calculated according to the CLPT method. All simulated correlation functions have been measured by employing the \nbodykit code, using $60$ linearly spaced bins in the range $1$--$121\Mpc$ for the bump cosmologies at $k_T=0.05\iMpc$, $30$ bins in the range $1$--$61\Mpc$ for models at $k_T=0.01$, and $15$ bins in the range $1$--$31\Mpc$ for models at $k_T=1.0\iMpc$. 

Fig.~\ref{corr-k0p05} shows the response functions for the bump at $k_T=0.05\iMpc$ for four different redshifts $z=2$, $1$, $0.5$, and $0$ and the bump cosmologies with widths $\sigma=1$, $0.3$, and $0.1$. We find a dip in the response around $85\Mpc$, which is due to a long wavelength modulation of the whole correlation function because of the localized bump feature in the corresponding power spectrum. By considering  equation~\eqref{linear-ps-bumpmodel}, the linear correlation function response becomes   
\begin{equation}
R_L(r) = 1 +  \frac{1}{\xi^\text{$\Lambda$CDM}_L(r)} \int_0^\infty d^3r'  \xi^\text{$\Lambda$CDM}_L(r') \tilde{F}(|\ve r- \ve r'|),    
\end{equation}
with
\begin{equation}
 \tilde{F}(r) = \int_0^\infty \frac{dk}{2 \pi^2} k^2  F(k) j_0(k r), 
\end{equation}
where $j_0$ is the zero-order spherical Bessel function. Hence for scales $r\gg k_T^{-1}$ we have that $ \tilde{F}(r) \rightarrow 0$ because of large cancellations provided by $j_0(k r)$; while as $r \rightarrow 0$,  $ \tilde{F}(r) \rightarrow (2 \pi^2)^{-1}\int dk k^2  F(k)$, a constant that when computed turns out to be much smaller than the correlation function at those small scales; such that the linear response goes to unity at both large scales and small scales limits. On the other hand, $k^2 F(k)$ reaches its maximum at scales $k_{max} > k_T$, the larger the width $\sigma$, the bigger are $k_{max}$ and the maximum values $k_{max}^2 F(k_{max})$; however for larger widths, the function  $k^2 F(k)$ overlaps with more oscillations and $\tilde{F}(r)$ is suppressed. The inverse transformed bump feature, $\tilde{F}(r)$, turns out to be a sinc-like function in real space with amplitude proportional to the $\sigma$ and first trough around $r_T \lesssim  (3/2) \pi/k_T $. Moreover, the aforementioned effects compete and this becomes more pronounced for the width $\sigma=0.3$ case. Coincidentally, for the case $k_T=0.05$, this trough is located at about the same scale as the dip, at about $87 \Mpc$, in the correlation function, and both reinforce each other to produce the large dip observed in Fig.~\ref{corr-k0p05}.

More generally, for the different choices of $k_T$ the changes in the correlation-function response will follow a similar pattern, but the wave-length modulations will be related to the characteristic scales: For larger $k_T$ and smaller $\sigma$, the oscillations in the correlation function have higher frequency and are damped at smaller scales. This can be seen in Figs.~\ref{corr-k0p1} and \ref{corr-k1}, for the cases $k_T=0.1$ and $1\iMpc$, respectively. The main qualitative difference compared to the $k_T=0.05\iMpc$ case is that the reinforcement with the BAO characteristic dip is not present in these cases.     

The peaks and troughs are more pronounced in the linear theory since the correlation function scales simply with the scale-independent linear growth function. On the other hand for CLPT, even in the Zeldovich approximation, overdense regions are partially depleted, while underdense regions populated, due to the free-streaming of coherent matter flows over a scale settled by the Lagrangian displacements 1-dimensional variance $\sigma^2_\Psi = \int_0^\infty dk P_L(k) / (6 \pi^2)$.\footnote{This effect has the same origin as the smearing of the BAO peak observed in LPT and simulations (see, e.g. \citealt{Tassev:2013rta}).} This effect is captured very well by our numerical results, and the damping in the responses become very similar for simulated data and CLPT. More remarkably for the cases of $k_T =1 \iMpc$ with $\sigma = 0.3$ and $0.1$, corresponding to middle and right panels of fig.~\ref{corr-k1}, where the oscillations in the power spectrum are practically erased. The reason for this is, of course, that the peaks and troughs are more closer to each other and the free-streaming of particles can cover such distances. Indeed, particles travel, on average, a distance settled by the standard deviation of the displacement field $\sigma_\Psi(z) \sim  6 \times D_+(z) \Mpc  $ with $D_+(z)$ the linear growth function, so the process of particles moving out from more dense regions becomes very efficient.


\section{Conclusions}\label{summary-conclusions}

Phase transitions in the dark sector are common in theories of cosmology beyond $\Lambda$CDM, and these leave fingerprints that are potentially detectable by current and future surveys. One of these signatures is the creation of enhanced features in the power spectrum at scales where otherwise the power would be smooth. The generation of these can be understood since all $k$-modes entering the horizon during the time elapsed by the phase transition suffer an enhancement on their amplitude since adding an extra $\rex$ increase the growth rate of the linear matter perturbations impacting modes $k\geq k_c$ entering the horizon for $a< a_c$. 
In this work we have focused on bumps generated in the power spectrum, motivated by the recently proposed BDE model of \cite{delaMacorra:2018zbk} and SEOS \citet{Jaber:2019opg}. 

We have studied non-linear evolution of parametric {\it bump cosmologies}. We have chosen to be as model independent as possible, instead of considering bumps generated by any specific BDE model, since we are interested in a wider range of theoretical models. We have also fixed the background cosmology to be \LCDM to allow us to investigate the phenomenology of bumps in isolation. In order to do so, we have run modest-resolution \nbody simulations, which are complemented by perturbation theory models and non-linear halo-model calculations from the \hmcode model of \cite{Mead2015}. We expect the different methods used to work over different ranges of scales, and this complementarity is important, since although bumps can be localized at a given scale, these are naturally spread by non-linear evolution, typically covering scales that may be outside the range of validity of some particular method. Bearing in mind that non-linear \LCDM is well studied, we have put attention to the power spectrum response, constructed as the ratio of the power in a bump cosmology to a cosmology with no bump, instead than on the power spectrum itself. Once an accurate model for the response is at hand, this can be converted into an accurate model for the power spectrum by multiplying by an accurate model for the \LCDM non-linear power spectrum. We have studied the non-linearities in both real and redshift space for the power spectrum and how these fingerprints are translated to configuration space in  the correlation function. 

Much of the non-linear physics is understood within the \hmcode method in the real space power spectrum. Of particular importance is the appearance of a second bump feature in the response generated at smaller scales than the first, primordial bump. The reason for this is a non-linear coupling of the bump and one-halo term in the following simple mechanism: long wave-length density fluctuations are enhanced to form the bump, but at the same time, small-scale fluctuations in regions located inside these overdensities, corresponding to 1-halo regions in halo models, are further amplified and can cross-over the threshold density for collapse more easily, leading to a more efficient halo formation than in a model without a bump. PT, on the other hand successfully follow the data at quasilinear scales, though it fails to model the second, non-linear bump, which is out of its reach. In redshift space this second, non-linear bump is partially erased because of the damping along the line-of-sight direction that is provided by the random motion of virialized regions that generate the "Fingers-of-God". Such effect is more clear in the quadrupole, for which the second bump is almost completely erased, since this multipole gives more weight to the line-of-sight direction. The monopole, on the other hand, still shows the second bump since it gives equal weight to all directions. This redshift-space effect, being highly non-linear, is not captured by perturbation theory, however at quasi-linear scales the simulated data and theory predicted by using the two popular methods of \cite{Taruya:2010mx} and \cite{Scoccimarro:2004tg} behave similarly. 

A localized bump in the power spectrum corresponds to oscillations in the correlation function with amplitude proportional to its width and a frequency governed by its position. The effect is to modulate the response about unity: higher wavenumbers at which the bump is located translate to higher oscillation frequencies; and wider bumps enhance the modulation, but are also more rapidly damped. This basic picture is explained well within linear Eulerian theory. By moving to Lagrangian space, we find that the signatures in the correlation function become even more damped since coherent flows have a finite probability to leave overdense regions and populate underdense regions. This effect has the same origin as the smearing of the BAO peak, that is well captured in LPT; and in the bump cosmology is much more evident for large-$k$ located bumps with small widths, since in these cases linear theory shows up rapid oscillations, and the displacement fields sizes, typically given by their standard deviations, are large enough, such that particles find the time to deplete the overpopulated regions. 

In the future it would be interesting to investigate the phenomenology of the bump cosmology for different bump amplitudes, which is fixed at $0.15$ in this paper, somewhat arbitrarily. One could also investigate more physical examples of `bump' cosmology, such as that generated by the physical BDE model, where the background expansion is also changed relative to \LCDM and where the bump shape will not necessarily be Gaussian. Since this was our first investigation, in this paper we focused on modelling statistics of the matter field, which unfortunately are not direct observables. In future, it would be fruitful to consider how the statistics of biased tracers of the density, such as haloes or galaxies, were affected by bump cosmologies. Specifically, we suspect that there may be interesting signatures generated in the halo bias and that these could be understood using \cite{Press1974} arguments in a similar way to how we could explain the generation of the `second bump' feature in the matter power spectrum. Of course, one could also investigate higher-order statistics, or statistics of transformed versions of the density fields \citep[e.g.,][]{Simpson2011, White2016}. On the theory side, it may be interesting to see if Effective Field Theory (EFT) \cite{Baumann:2010tm} could be used to extend the reach of perturbation theory. Compared to SPT, EFT simply adds a term $-c_s^2 k^2 P_L(k)$ to the power spectrum, with fitted effective speed of sound $c_s^2$, and this extra freedom may allow for a joint EFT-halo model approach to accurately model the response across all scales. 

Summarizing, in this work we have used non-linear methods to study the fingerprints that may be left by cosmological models on which the dark energy suffers phase transitions, and have the potential to be detectable by current and future galaxy surveys.

\section*{Data availability}
\hmcode is publicly available at \hmcodelink. The perturbation theory code \textsc{mgpt} is publicly available at \href{https://github.com/cosmoinin/MGPT}{https://github.com/cosmoinin/MGPT}. The simulated data and theory curves generated as part of this work will be shared on reasonable request to the corresponding author.

\section*{Acknowledgements}
 DG and AM acknowledge partial support from Project IN103518 PAPIIT--UNAM,  DG thanks support from a CONACyT PhD fellowship and AM  from PASPA--DGAPA, UNAM and CONACyT. AJM has received funding from the Horizon 2020 research and innovation programme of the European Union under Marie Sk\l{}odowska-Curie grant agreements No. 702971. AA acknowledges partial support from Conacyt Grant No. 283151.

\label{lastpage}


\bibliographystyle{mnras}
\bibliography{bibliography} 

\end{document}